\documentclass{article}

\usepackage{graphicx}
\usepackage{hyperref}
\usepackage{amsmath}
\usepackage{amssymb}
\newcommand{\mv}{\mathbf}	     

\title{Predicting Spatio-Temporal Time Series Using Dimension Reduced Local States}
 
 \author{Jonas Isensee$^{1,2}$,  George Datseris$^{1,2}$ and Ulrich Parlitz$^{1,2}$ \\[3mm]
$^{1}$Max Planck Institute for Dynamics and Self-Organization, \\ Am Fassberg 17, 37077 G\"ottingen, Germany \\
$^{2}$Institute for Nonlinear Dynamics, Georg-August-Universit\"at G\"ottingen, \\ Friedrich-Hund-Platz 1, 37077 G\"ottingen, Germany}

\date{\today}

\begin{document}

\maketitle

\begin{abstract}
We present a method for both cross estimation and iterated time series prediction of
spatio temporal dynamics based on reconstructed local states, PCA dimension reduction, 
and local modelling using nearest neighbour methods.
The effectiveness  of this approach is shown for (noisy) data
from a (cubic) Barkley model, the 
Bueno-Orovio-Cherry-Fenton model, and
the Kuramoto-Sivashinsky model.
\end{abstract}

\section{Introduction}
In many experiments some variables of the system are more easily observable than others. If the underlying 
dynamics is deterministic, in general the observable of interest is nonlinearly related to other variables of the system 
which might be more accessible.
In such cases one may try to estimate any observable which is difficult to measure from time series of 
those variables which are at one's disposal. 
Another task frequently encountered with observed time series is forecasting the dynamical evolution of the system and the time series. To cope with both tasks in case of multivariate time series from extended spatio-temporal systems
we present an approach for cross estimation and iterated time series prediction
using local state reconstructions, dimension reduction, and nearest neighbour methods for local modelling.
Local state reconstruction is motivated by the fact that it often is
impractical to predict the behaviour of systems with a large spatial extent all at once.  
If instead one combines a spatial and temporal neighbourhood around each measurement to find
a description of the local system state it becomes possible to make predictions for
each point in space independently.
For performing cross estimation or prediction based on local states one can either use nearest neighbours methods (also called local modelling) 
 \cite{parlitz_prediction_2000}  or employ some other black-box modelling approach like, for example, Echo State Machines  \cite{pathak_model-free_2018,zimmermann_observing_2018}.
 In the following, we shall use local modelling by selecting for each reconstructed reference state similar states from a training data set whose relations to other observables and/or future temporal evolutions are known and can be exploited for cross estimation or time-series prediction.
 
Successful reconstruction of high-dimensional  dynamics in extended systems, however, requires very large embedding dimensions which is a major challenge in particular for nearest neighbour methods. 
Therefore, a crucial point in making the conceptually simple nearest neighbours algorithm performant is dimension reduction.
As a means of dimension reduction to find lower a dimensional representation of the local states, 
we employ  principal component analysis (PCA) which turns out to improve performance in particular for noisy data.

\section{Predicting Spatio-Temporal Time Series}

In this section we shall introduce the main concepts for predicting spatio-temporal time series, including local delay coordinates states, linear dimension reduction, and nearest neighbours methods for local modelling of the dynamical evolution or any other relation between observed time series.

\subsection{Local Modelling}
Let $\mathbf{x}_t$ be a state of some dynamical  system evolving in time $t$ and let $s_t = h ( \mathbf{x}_t )$ be a signal which can be observed or measured. Furthermore, let's assume that the dynamical equations generating the flow in state space and the measurement function $h$ are unknown, but only a set ${\cal{S}} $ of $M$ states $\mathbf{x}_{t_m}$ and corresponding time series values $s_{t_m}$ for $t_1, \hdots , t_M$ are available, for which also future values   $\mathbf{x}_{t_m+T}$  and $s_{t_m+T}$ are known (due to previous measurements, for example).  This data set ${\cal{S}} $ can be used to predict the future value  $\mathbf{x}_{t+T}$ of  a given state $\mathbf{x}_t$  or to estimate the corresponding time series values  $s_t$ and $s_{t+T}$, by selecting the $k$ nearest neighbours of $\mathbf{x}_t$ in  ${\cal{S}} $ and using  their future values (or the corresponding time series values) for approximating $\mathbf{x}_{t+T}$ (or  $s_t$ and  $s_{t+T}$), for example,  by (distance weighted) averaging.

In most practical applications of this kind of local nearest neighbour modelling the required states are reconstructed from a measured time series using the concept of delay coordinates (to be introduced in the next section).
Local modelling in (reconstructed) state space is a powerful tool for purely data driven  time series prediction \cite{Atkeson1997,engster_local_2006}. Its main ingredients are a proper state space representation of the measured time series, fast nearest neighbour searches, and  local models such as low order polynomials which can accurately interpolate 
and predict the (nonlinear) relation between (reconstructed) states and target values.

\subsection{Delay Coordinates}
The most important part of time series based  local modelling 
 is the representation of data, i.e. proper reconstruction of states from data.
Typically this representation is found utilizing delay coordinates and  Taken's Embedding Theorem 
\cite{takens_detecting_1981, sauer_embedology_1991,kantz_nonlinear_2004,Abarbanel_1996,BK_2015}
such that a scalar time series $\{s_t \}$ is reconstructed to state vectors
\begin{equation*}
    \mv x_t= (s_{t-\gamma\tau}, \dots,  s_{t-\tau}, s_{t})
\end{equation*}
by including $\gamma$ past measurements each separated by $\tau$ time steps.
For multivariate time series $\{\mathbf{s}_t\}$ one can do the same for each of the components resulting is state vectors 
\begin{equation*}
    \mv x_t= (s_{1,t-\gamma\tau}, \dots, s_{1,t},s_{2,t-\gamma\tau}, \dots, s_{2,t}).
\end{equation*}

\subsection{Spatial Embedding} \label{sec:st_emb}
In principle, delay embedding could also be employed to reconstruct (global) states of
high-dimensional spatially extended systems using multivariate time series sampled at many spatial locations.
Such global state vectors are (and have to be) very high dimensional (for systems exhibiting extensive chaos). 
The runtime of nearest neighbour searches, however, and in particularly the memory usage of such reconstructions
grows rapidly with the dimension of the reconstructed global states.
To avoid this issue it has been  proposed  \cite{Parlitz_98,parlitz_prediction_2000,MGG_2004} to reconstruct 
(spatially) local states and  to predict  spatially extended systems point by point instead of the whole global state at once.
This approach is motivated by the fact that 
all spatially extended physical systems posses a finite speed at which information travels.
Therefore the future value of any of the variables depends solely on its past and
its spatial neighbours.\footnote{The situation is different, if additional long-range connections exist linking remote locations.}
Instead of trying to reconstruct the state of the whole
system into one vector, we limit ourselves to reconstructing small neighbourhoods of all
points that carry enough information to predict one point one time step into the future.
As an additional benefit the infeasibly large embedding dimension that would result from 
embedding the entire space into a single state is greatly reduced.
The idea of local state reconstruction was first applied to spatially one-dimensional systems 
\cite{Parlitz_98,parlitz_prediction_2000,MGG_2004} and was used, for example,
to anticipate extreme events in extended excitable systems \cite{Bialonski}.

In the following we will present the embedding procedure for spatiotemporal time series
represented by $u_{t,\alpha}$, where $t$ denotes time and $\alpha$ a point in space.
For 2D space $\alpha$ takes the values $\alpha = (i,j)\quad \forall 1 \leq i \leq N_x,\,1 \leq j \leq N_y $.

\begin{figure}[h!]
    \centering
    \includegraphics[width=0.9\textwidth,keepaspectratio=true]{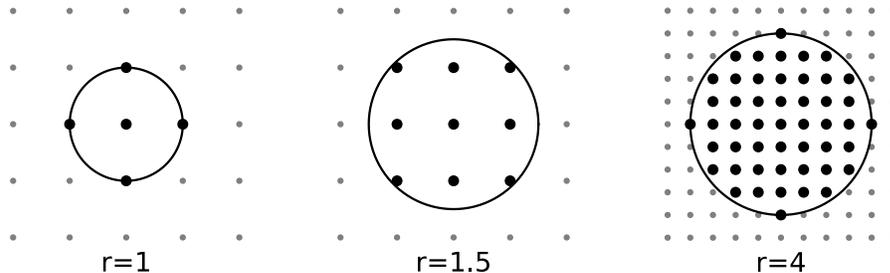}
     \caption{Visualization of spatial embedding for radii $r \in \{1,\, 1.5, \, 4\}$. All points
    within the circle spanned by $r$ are included in the embedding.}
    \label{fig:embedding}
\end{figure}

In the most general case such an embedding could consist of arbitrary combinations
of neighbours in all directions of space and time.
For practical purposes we will limit ourselves to a certain set of parameters to
describe which neighbours will be included into a reconstruction.
We parameterize an embedding with the number $\gamma$ of past time steps and
their respective temporal delay (or time lag) $\tau$.
All neighbours in space that are within the radius $r$, referring to the 
Euclidean distances in a unit grid, will be included as well. 
The resulting shape of the embedding is comparable to a cylinder in 2+1D space-time.
To make this clearer, a
visualization of the spatial embedding in a two-dimensional system is displayed in Fig.~\ref{fig:embedding} for different radii $r$.

In the following we shall assume that the dynamics underlying the observed spatio-temporal time series 
is invariant with respect to translations, i.e. that the system is homogeneous. In this case states reconstructed 
at different locations can be combined to a single training set providing the data base for cross estimation or time series prediction as will be discussed in more detail in Section \ref{sec:predalg}. However, even if the dynamical rules are the same 
for all locations, special care needs to be taken at the boundaries. This becomes obvious when trying to
include nonexistent neighbours from outside the grid. For periodic boundary condition
the canonical solution is to wrap around at the edges but for constant boundaries
the solution is not so obvious.
In many cases the dynamics
near the boundary may also differ from dynamics far from it.
It is therefore desirable to treat boundaries separately during
nearest neighbour predictions.
A solution proposed in Ref.~\cite{parlitz_prediction_2000} is to fill the missing
values at the boundary with an additional parameter being a large constant number.
If the parameter is significantly larger
than typical values of the internal dynamics, the states reconstructed from the boundary
fill regions in state space isolated from state vectors of internal dynamics.
This has the desired effect as nearest neighbour searches will always find
boundary states when given a boundary state as query and similarly for internal states.

\subsection{Dimension Reduction} \label{sec:predalg}
The feasibility of any nearest neighbour search depends heavily on the memory consumption
as the whole data set needs to lie in memory. A crucial part of our algorithm is therefore about creating a proper 
low dimensional reconstruction. In most numerical experiments choosing just very few neighbours to include 
in the reconstruction did not prove to be very effective. Therefore, instead of choosing a small embedding dimension from the start, we propose to perform some means of dimension reduction on the resulting reconstructed state vectors. For this task we choose Principal Component Analysis (PCA) as it is a straight-forward  standard technique for (linear) dimension reduction, where the reconstructed states $\mathbf{x}_t$ are projected onto the eigenvectors of the covariance matrix corresponding the largest eigenvalues \cite{Gareth_2015}. In the field of nonlinear time series analysis PCA has first been used by D. Broomhead and G. King \cite{BK_1986}  who suggested to use dimension reduction applied to high dimensional delay reconstructions with time series densely sampled in time.


Let $\{  \mathbf{x}^n\}$ be the set of all $N$ reconstructed states  
$ \mathbf{x}^n = (x_1^n, \hdots, x_{D_E}^n)\in \mathbb{R}^{D_E}$
(at different times $t$ and locations $\alpha$, assuming stationary and spatially homogeneous dynamical rules).
To perform PCA first mean values
$\mathbf{\bar x} =   \frac{1}{N} \sum_{n=1}^N   \mathbf{x}_n= (\bar x_1, \hdots, \bar x_{D_E})$  with 
$\bar x_i = \frac{1}{N} \sum_{n=1}^N x_i^n$ are subtracted resulting in shifted states 
$\mathbf{\tilde x}^n= \mathbf{x}^n- \mathbf{\bar x} = (\tilde x_1^n, \hdots , \tilde x_{D_E}^n)$. 
 The covariance matrix 
\begin{eqnarray*}
   C_{X} & = & \frac{1}{N} \sum_{n=1}^N (\mathbf{\tilde x}^n)^{tr} \cdot \mathbf{\tilde x}^n  \\
               & = & \frac{1}{N} 
                      \begin{bmatrix}
                        \sum_{n=1}^N  \tilde x_1^n  \tilde x_1^n    & 
                        \sum_{n=1}^N  \tilde x_1^n  \tilde x_2^n    & 
                        \hdots & 
                        \sum_{n=1}^N  \tilde x_1^n  \tilde x_{D_E}^n     \\
                        \sum_{n=1}^N  \tilde x_2^n  \tilde x_1^n    &
                        \sum_{n=1}^N  \tilde x_2^n  \tilde x_2^n    & 
                         \dots & 
                         \sum_{n=1}^N   \tilde x_2^n   \tilde x_{D_E}^n  \\
                        \vdots &
                        \vdots & 
                        \ddots & 
                        \vdots \\
                         \sum_{n=1}^N \tilde x_{D_E}^n   \tilde x_1^n   & 
                         \sum_{n=1}^N \tilde x_{D_E}^n   \tilde x_2^n  & 
                         \hdots & 
                         \sum_{n=1}^N   \tilde x_{D_E}^n  \tilde x_{D_E}^n  \\
                         \end{bmatrix}
 \end{eqnarray*}

is approximated by iteratively reconstructing individual shifted delay vectors $\mathbf{\tilde  x}^n$ from the dataset 
and summing the terms $(\mathbf{\tilde x}^n)^{tr} \cdot \mathbf{\tilde x}^{n}$ into the preallocated matrix $C_{X}$.

Local states $\mathbf{y}^n$ of lower dimension $D_R \le D_E$ are obtained by projecting the shifted states $\mathbf{\tilde x}$
\begin{equation*}
  \mathbf{y}^n = P \mathbf{ \tilde x}^n 
\end{equation*}
using a (globally valid) $D_R \times D_E$ projection matrix $P$ whose rows are given by the 
$D_R$ eigenvectors of the matrix $C_{X}$ corresponding to the largest $D_R$ eigenvalues.
The dimensionality $D_R$ of the subspace spanned by  eigenvectors to be taken into account can either be set explicitly or
determined such that some percentage of the original variance of the embedding is preserved.

The whole data set can thus be embedded into the space with reduced dimension $D_R$ by 
embedding each point of the data set into the high dimensional space $\mathbb{R}^{D_E}$ and projecting it into
the low dimensional space $\mathbb{R}^{D_R}$ using the PCA projection matrix $P$ computed beforehand.

For the subsequent prediction process the projected reconstruction vectors $\mathbf{y}^n$ are then fed into a tree structure such as a kd-tree \cite{bentley_multidimensional_1975, carlsson_nearestneighbors.jl:_2018} for fast
nearest neighbour searching.

One issue arises with states near boundaries. Since the dynamics close to the
boundaries may differ from the rest of the system, they were separated from
other reconstructed vectors in phase space. This was achieved by setting the non-existent
neighbours of boundary points to a large constant value \cite{parlitz_prediction_2000}.
The power of PCA however relies on its assumption of a single cloud of points  in (state) space close 
to a low dimensional linear subspace. This is no longer the case when constant boundaries come into play.
To sidestep this issue we suggest changing the second step of the procedure described above.
Simply exclude all boundary states from the computation of the projection matrix $P$ but project them 
with the resulting matrix $P$ nonetheless.
In principle this could eliminate the offset meant to separate internal and boundary dynamics
but in practice the projection matrices rarely posses zero-valued entries. Therefore
it is highly unlikely that this would become a problem as long as boundary offset values
are chosen large enough.

\subsection{Prediction Algorithm}
While the values and the dimension of the state vectors have changed in
the dimension reduction process, their ordering $(t,\alpha)  \leftrightarrow n $ within
the reconstructed space and the search tree is known.
It is therefore sufficient to find the indices of nearest neighbours in the dimension reduced 
reconstructed training set in $\mathbb{R}^{D_R}$.
To make predictions we assign each reconstructed state $\mv{x}_{t,\alpha}$ a target value $\mathbf{z}_{t,\alpha}$ 
from the original training data and the only difference between temporal prediction and cross estimation lies in the choice of
these target values.

For time series prediction we choose
$\mv{x}_{t,\alpha} \rightarrow u_{t+1,\alpha}$ where $\mv{x}_{t,\alpha}$ are the
reconstructed vectors from the spatio-temporal time series $\{ u_{t,\alpha} \}$ and $u_{t+1,\alpha}$
the target values.
The prediction process then consists of reconstructing $\mv{x}_{T,\alpha}$ from the end of the time series by
applying the same embedding parameters, subsequent dimension reduction using the projection matrix $P$ 
that was computed for the training set, and local nearest neighbour modelling providing the target values $u_{T+1,\alpha}$. 
Once a prediction for each point (denoted by $\alpha$) has been made, all future values $u_{T+1,\alpha}$ of the 
(input) field $u$ are known and the procedure can be repeated for predicting $u_{T+2,\alpha}$. Using this kind of iterated 
prediction spatio-temporal time series can, in principle,  be forecasted for any period of time (with the well-known limits of predictability of chaotic dynamics). 

The case of cross estimation is even simpler than time series prediction. Here
we are given a training set of two fields:
an input variable $u_{t,\alpha}$ and a target variable $v_{t,\alpha}$.  The values of the input field
$u_{t,\alpha}$ are reconstructed into delay vectors $\mv{x}_{t,\alpha}$. 
Using PCA and nearest neighbours search we find similar reconstructed states in the training set
for which the corresponding values of the target variables are known and can be used for estimating the current 
target  $v_{t,\alpha}$.

\subsection{Error Measures}
In the next section we will test the presented prediction methods on the model systems
described in section~\ref{sec:model_systems}. For evaluation we compare any predicted field $\hat v$ with the corresponding 
correct values (i.e. test values) $\check v$ by considering spatial averages over all sites $\alpha$. This so-called Mean Squared Error (MSE) is then normalized by the MSE obtained when using the (spatial) mean value $\bar v$ for prediction. The resulting 
Normalized Mean Squared Error ($\text{NRMSE}$) is defined as
\begin{align}
  \text{\text{NRMSE}}(\check v,\hat v) =
  \sqrt{\frac{\text{\text{MSE}}(\check v,\hat v)}{\text{\text{MSE}}(\check v,\bar v)}},
  \label{eq:nrmse} \hspace{0.5cm}\text{where}\hspace{0.5cm}
  \text{\text{MSE}}(\check v,\hat v) = \frac{1}{A}\sum_{\alpha}
  \left(\check v_{\alpha} - \hat v_{\alpha}\right)^2
\end{align}
where  $A$ is the number of spatial sites $\alpha$ taken into account. Any good estimate or forecast should be 
(much) better than the trivial prediction using mean values and result in NMSE values (much) smaller than one. 

\subsection{Software}
All software used in this paper has been published in the form of an 
open source software
library under the name of \emph{TimeseriesPrediction.jl}~\cite{timeseriesprediction_2018}
along with extensive documentation and various examples.
It is written using the programming language Julia~\cite{bezanson_julia:_2017}
with extensibility in mind, such that it is compatible
with different spatial dimensions as well as arbitrary spatiotemporal embeddings.
This is made possible through a modular design and Julia's multiple dispatch.

\section{Model Systems}  \label{sec:model_systems}

The Kuramoto-Sivashinsky (KS) model
\cite{kuramoto_diffusion-induced_1978, sivashinsky_flame_1980, sivashinsky_nonlinear_1988}
has been devised for modelling flame fronts and will
in our case be used as a benchmark system for iterated time series prediction.
The Barkley model \cite{barkley_model_1991} describes an excitable medium that shows
 chaotic interplay of traveling waves.
The third and most complex model is the Bueno-Orovio-Cherry-Fenton (BOCF) model
\cite{bueno-orovio_minimal_2008}, which is composed of four coupled fields describing electrical excitation
waves in the heart muscle.

\subsection{Kuramoto-Sivashinsky System}
The Kuramoto-Sivashinsky (KS) system
\cite{kuramoto_diffusion-induced_1978, sivashinsky_flame_1980, sivashinsky_nonlinear_1988}
is defined by the following partial differential equation:
\begin{align}
  \frac{\partial u}{\partial t} + \frac{\partial^2u}{\partial x^2} + \frac{\partial^4 u}{\partial x^4} + \left|\frac{\partial u}{\partial x}\right|^2 = 0
   \label{eq:ks}
\end{align}
typically integrated with periodic boundary conditions.
It is widely used in literature \cite{parlitz_prediction_2000,pathak_model-free_2018}
because it is a simple system consisting of just one field while still showing highdimensional chaotic dynamics. 

\begin{figure}[h!]
  \centering
  \includegraphics[width=.85\textwidth,keepaspectratio=true]    {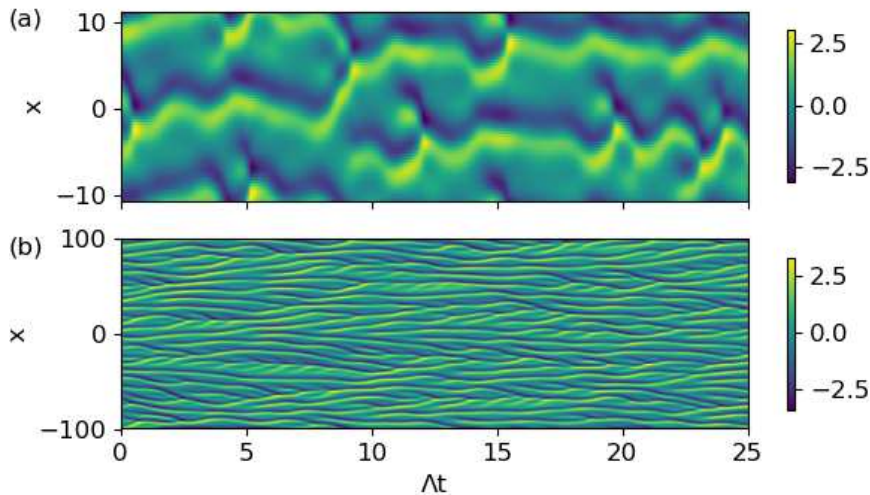}
  \caption{Temporal evolution of the KS model (\ref{eq:ks}) for two different system sizes. Pane (a)
  has parameters $L=22$ and $Q=64$, while the larger system (b) has $L=200$ and  $Q=512$.}
  \label{fig:ks_ex}
\end{figure}

The dynamics were simulated with an EDTRK4 algorithm~\cite{cvitanovic_chaos_2016, kassam_fourth-order_2005}
and the parameters for integration
are the time step $\Delta t=0.25$ and the system size $L$ with spatial sampling $Q$.
Two example evolutions with $L=22$, $Q=64$ and $L=200$, $Q=512$ are shown in
Fig.~\ref{fig:ks_ex}.

\subsection{Barkley Model}
\label{sec:bk_def}
The Barkley model \cite{barkley_model_1991} is a simple system that exhibits excitable dynamics.
We will use a modification with a cubic $u^3$ term that can be used to
generate spatio-temporal chaos such that:
\begin{align}
  \begin{aligned}
    \frac{\partial u}{\partial t} =&\, \frac{1}{\varepsilon}u(1-u)\left(u-\frac{v+b}{a}\right)
    + D\nabla^2u \\
    \frac{\partial u}{\partial t} =&\, u^3 - v,
  \end{aligned} \label{eq:bk}
\end{align}
where the parameter set $a=0.75$, $b=0.06$, $\varepsilon=0.08$ and $D=0.02$
leads to chaotic behavior. For integration we used $\Delta t = 0.01$
and $\Delta x =0.1$ in combination with an optimized FTCS scheme like the one
described in \cite{barkley_model_1991}.

\begin{figure}[h!]
  \centering
  \includegraphics[width=.85\textwidth,keepaspectratio=true]
  {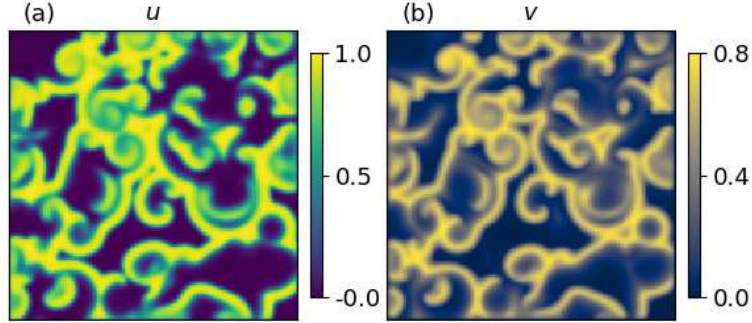}
  \caption{Snapshot of the chaotic Barkley model (\ref{eq:bk}) on a grid of size $150\times150$ with constant
  boundary conditions and after transients decayed.   The $u$ variable is displayed in (a) and $v$ in (b).}
  \label{fig:bk_ex}
\end{figure}

\subsection{Bueno-Orovio-Cherry-Fenton Model}
\begin{figure}[h!]
  \centering
  \includegraphics[width=\textwidth,keepaspectratio=true]{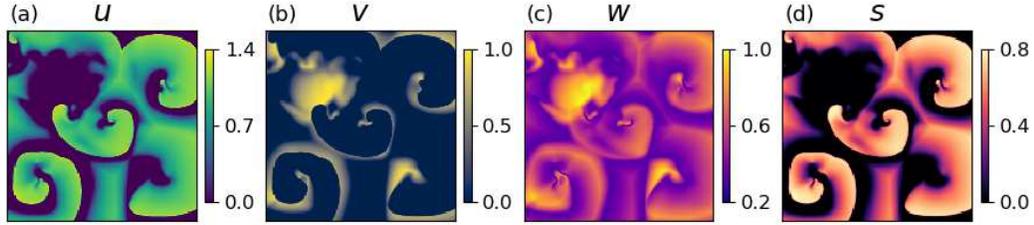}
  \caption{Snapshot of the BOCF model simulated on a $500\times500$ grid and
  coarse grained to a $150\times150$ grid using the software by Zimmermann \cite{zimmermann_observing_2018}.}
  \label{fig:bocf_ex}
\end{figure}

The Buono-Orovio-Cherry-Fenton (BOCF) model \cite{bueno-orovio_minimal_2008}
is a more advanced set of equations that serves as a realistic but relatively simple model of (chaotic)
cardiac dynamics. It consist of four coupled fields
that can be integrated as PDEs on various geometries.
For the sake of simplicity we consider a two-dimensional square.
The four variables $u$, $v$, $w$, $s$ are given by the following equations:
\begin{align}
  \begin{aligned}
  \frac{\partial u}{\partial t} =&
      D\cdot \nabla^2u-(J_{si}+J_{fi}+J_{so}) \\
  \frac{\partial v}{\partial t} =&
      \frac{1}{\tau_v^-}(1-H(u-\theta_v))(v_\infty -v)-\frac{1}{\tau_v^+}H(u-\theta_v)v\\
  \frac{\partial w}{\partial t} =&
      \frac{1}{\tau_w^-}(1-H(u-\theta_w))(w_\infty-w)-\frac{1}{\tau_w^+}H(u-\theta_w)w\\
  \frac{\partial s}{\partial t} =&
      \frac{1}{2\tau_s}((1+\tanh(k_s(u-u_s)))-2s)
  \end{aligned}
  \label{eq:bocf}
\end{align}
 where the currents $J_{si}$, $J_{fi}$ and $J_{so}$ and all other parameters
are defined in the appendix.
Variable $u$ represents the voltage across the cell membrane and provides 
spatial coupling due the diffusion term, whereas $v$, $w$, and $s$ are governed by local ODEs without any spatial coupling.
Fig.~\ref{fig:bocf_ex} shows a snapshot of all four fields. To make it easier to tell the
different fields apart each one has been assigned its own color map that will be used
consistently.
For simulation we used an implementation by Roland Zimmermann \cite{zimmermann_observing_2018},
 that simulates the dynamics of the BOCF model using an FTCS scheme on a $500\times500$ grid with
integration parameters $\Delta x = 1$, $\Delta t = 0.1$, diffusion constant $D=0.2$, 
no-flux boundary conditions and a temporal sampling of $t_{\text{sample}} = 2.0 $.
The dense spatial sampling is needed for integration but impractical for our use.
Therefore the software by Zimmermann coarse-grains the data to a grid of size $150\times150$.

\section{Cross Estimation}  \label{sec:cross_est}

For cross estimation we analyze the Barkley model and the BOCF model.
In the beginning both systems are simulated for more than $10000$ time steps so that different
subsets can be chosen for model training and testing. All training sets consisted
of 5000 consecutive time steps. Due to the dense temporal sampling the
first few frames after the end of the training set are potentially predicted much better than the following ones, because data points nearly identical to the desired estimation  output are already included in the training set. To avoid this predictions were offset by 1000 frames after
the end of the training sequence and averaged over 20 predicted frames
to reduce fluctuations and compute a standard deviation for the error measures.

\begin{figure}[h!]
  \centering
  \includegraphics[width=\textwidth,keepaspectratio=true]{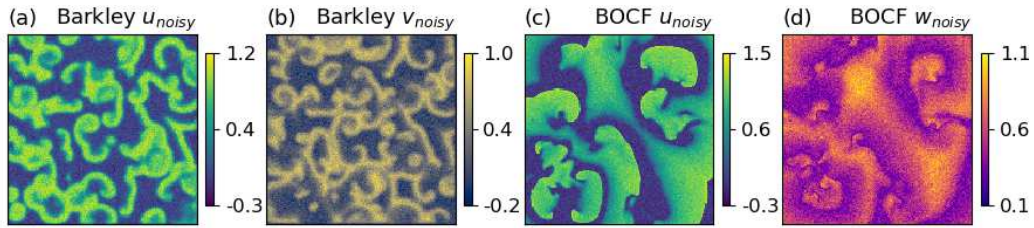}
  \caption{Snapshots of the variables $u$ and $v$ of the Barkley model and the variable $u$ and $w$ of the BOCF model
  after addition of normally distributed noise.}
  \label{fig:noisy_input}
\end{figure}

To simulate uncertainty in measurements normally distributed random numbers
were added to the observed variable in the test set.
Adding such noise with mean $\mu=0$ and standard deviation  $\sigma_\mathcal{N}=0.075$ resulted in signal-to-noise ratios
\[\text{SNR}_{\text{db}} = 10\log_{10}
 \frac{\left<u_{t,\alpha}^2\right>}{\sigma_{\mathcal{N}}^2}\]
 of 18.5dB and 14.6dB for $u$ and $v$ in the Barkley model, respectively. For the fields $u,v,w,s$ of the BOCF model SNRs were 20.1dB, 13.2dB, 18.1dB, and 15.4dB, respectively. For an intuition of the noisyness Fig.~\ref{fig:noisy_input} shows the variable $u$ and $v$ of
the Barkley model and the variables $u$ and $w$ of the BOCF model with added noise.

\subsection{Barkley Model}
\label{sec:bk_cross}
For the Barkley model (\ref{eq:bk}) only the $u$ variable has a diffusion term. Therefore the dynamics of $v$
solely depends on $u$ and its past. This significantly reduces the reconstruction parameter space as
spatial neighbourhoods may only be needed for noise reduction during PCA and can 
likely be small. For the prediction direction $u\rightarrow v$ the best embedding found by
optimization, the one with least prediction error, was $\gamma=500$, $\tau=1$ and $r=0$.
These parameters produce a highly redundant embedding which allows PCA
to efficiently filter out noise.
The other direction $v\rightarrow u$ needs spatial neighbourhoods for effective cross prediction
and the embedding parameters were $\gamma=30$, $\tau=5$ and $r=3$.

\begin{figure}
  \centering
  \includegraphics[width=\textwidth,keepaspectratio=true]  {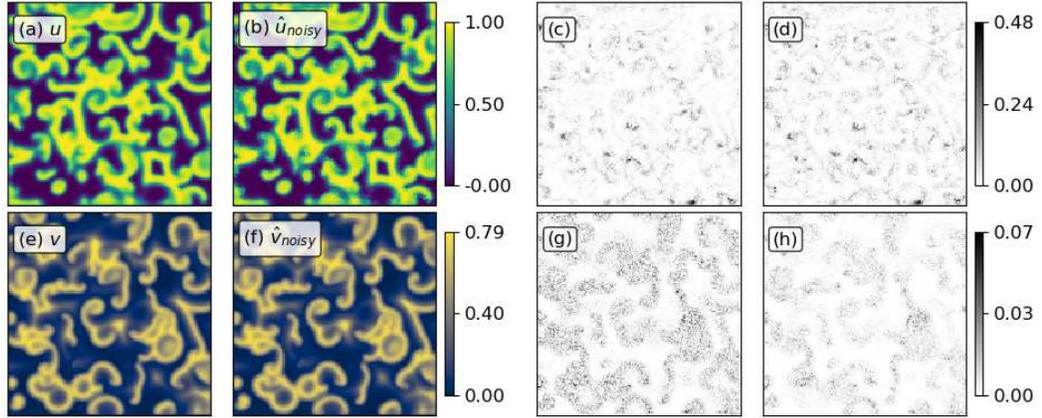}
  \caption{Cross estimation of data generated by the Barkley model (\ref{eq:bk}) 
  from a noisy $v$ field to $u$ and vice versa.   (a)-(d) show estimates of the $u$ field where (a)
  is the actual $u$ field, (b) the predicted field $\hat u$ (from noisy input), (c) the absolute difference
  between the two, and (d) a reference estimation error for noiseless input with 
  identical embedding parameters and training set.
  Panes (e)-(h)  show the same for the field $v$.
  The embedding parameters are listed in Table~\ref{tab:bk_errs}.}

  \label{fig:bk_cp}
\end{figure}

The results evaluated according to the error measure (\ref{eq:nrmse}) 
are listed in Table~\ref{tab:bk_errs}. A visualization of the predictions is shown in Fig.~\ref{fig:bk_cp} 
along with additional predictions performed with identical parameters  but for noiseless input.

\begin{table}
\center
      \begin{tabular}{@{}lll}
  \hline
    &
    $u\rightarrow v$ & $v\rightarrow u$\\
   \hline
    $\gamma$         &  500  &  30 \\
    $\tau$      &   1   &  5 \\
    $r$         &   0   &  3  \\
    $D_E$       &   501 & 899\\
    $D_R$       &   15   &  15 \\
   \hline
    $\text{NRMSE}$       & $0.0354\pm0.0013$ & $0.096\pm0.006$ \\
   \hline
    \end{tabular}
    \caption{\label{tab:bk_errs}Optimal embedding parameters and cross estimation errors for noisy data from the 
    Barkley model (\ref{eq:bk}) with temporal sampling $t_{\text{sample}}=0.01$. 
    $D_E$ is the initial embedding dimension (a direct result of the reconstruction parameters chosen) and 
    $D_R$ is the reduced dimension used to make the prediction. For both predictions we used the constant value of 200 for the      
     beyond the boundary pixels. The errors are averaged over 20 predicted frames.}    
\end{table}

\subsection{BOCF Model}
Similarly to the Barkley model only the $u$ variable of the BOCF model (\ref{eq:bocf}) has a diffusion term
which simplifies the predictions of $u\rightarrow\{v,w,s\}$. 
All embedding parameters, found with a stochastic gradient decent procedure, are listed
along with the prediction errors in Table~\ref{tab:bocf_errs}.
In most of these cases we observe a very large temporal embedding with a small spatial neighbourhood. This is likely due to the dense temporal sampling relative to
the propagation speed of wavefronts within the simulated medium.
In this way the highly redundant embedding and PCA for dimension reduction
provide an effective method of noise reduction.
The $w$ field however presents itself as a somewhat smeared out version of the other variables
thus requiring a larger spatial neighbourhood to recover the positions of wavefronts.

\begin{figure}[ht!]
  \centering
  \includegraphics[width=\textwidth,keepaspectratio=true] {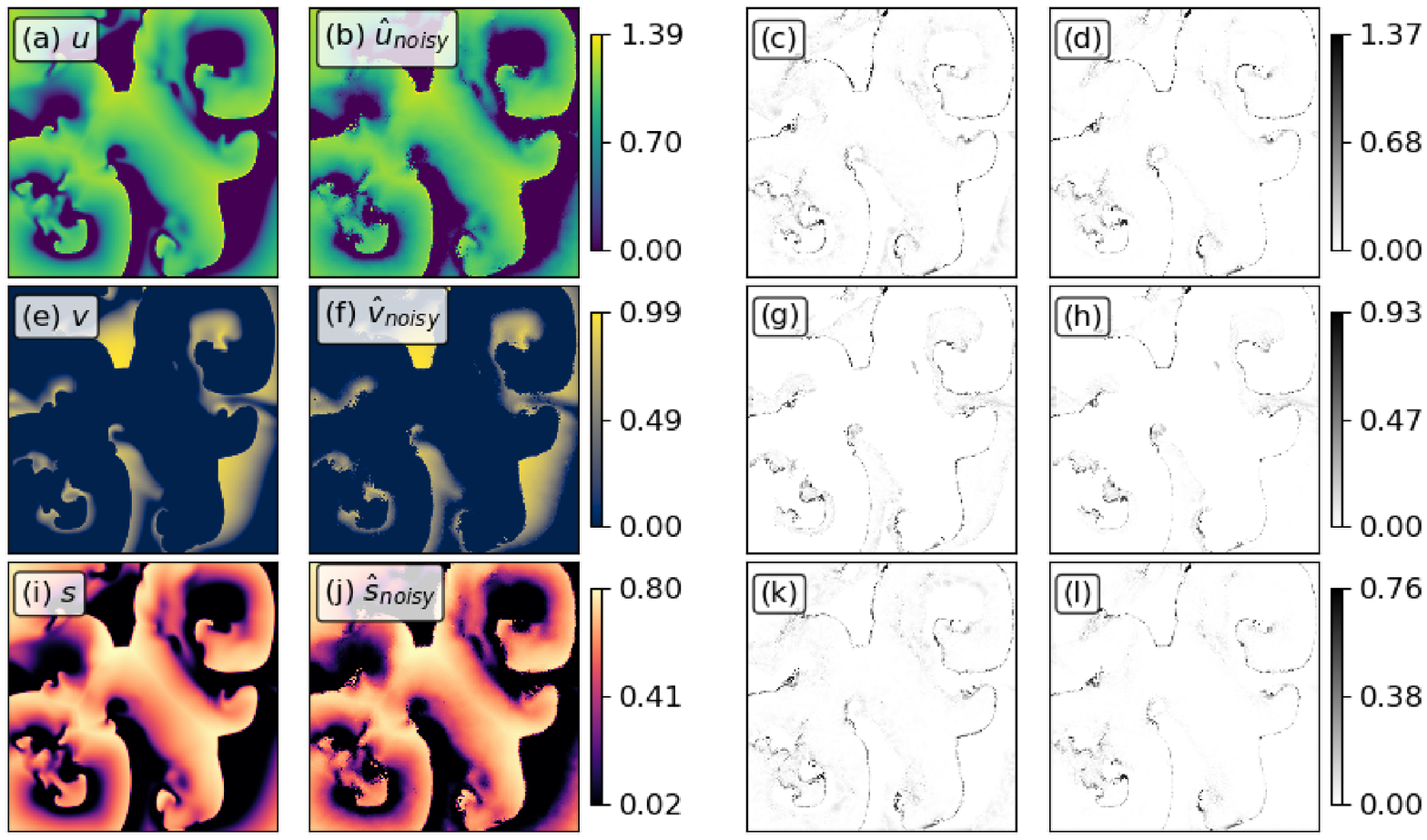}
  \caption{Cross estimation of data generated by the BOCF model  (\ref{eq:bocf}) from $w_{noisy}$ to all three
  other variables. (a)-(d) show estimates of the $u$ field where (a)
  is the actual $u$ field, (b) the predicted field $\hat u$ (from noisy input), (c) the absolute difference
  between the two, and (d) a reference estimation error for noiseless input with 
  identical embedding parameters and training set.
  Panes (e)-(h) and (i)-(l) show the same for their fields $v$ and $s$, respectively.
  The embedding parameters are listed in Table~\ref{tab:bocf_errs}.}
  \label{fig:bocf_w_pred}
\end{figure}

\begin{figure}[ht!]
  \centering
  \includegraphics[width=\textwidth,keepaspectratio=true]  {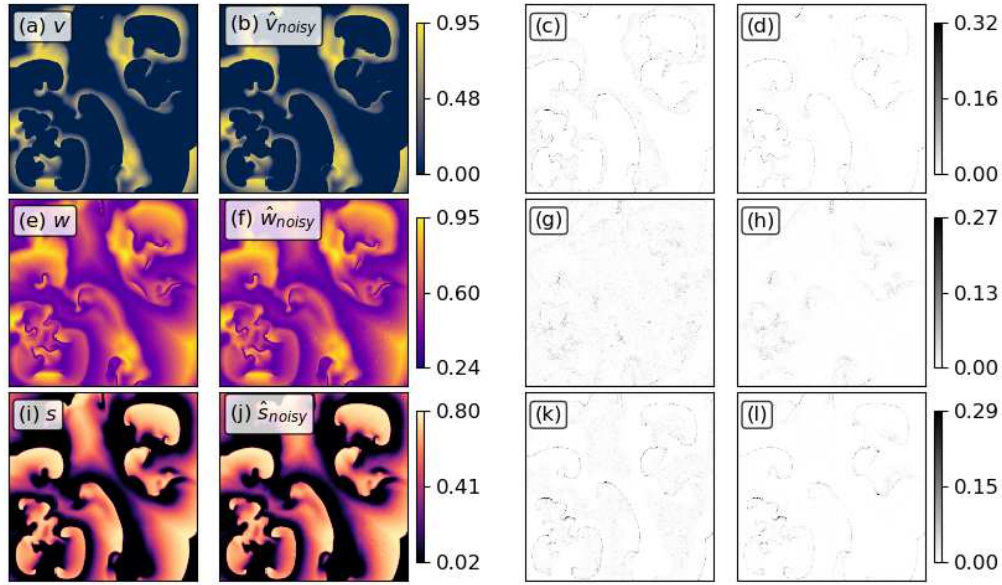}
   \caption{
Cross estimation of data generated by the BOCF model (\ref{eq:bocf}) from $u_{noisy}$ to all three
  other variables. (a)-(d) show estimations for the $v$ field where (a)
  is the actual $v$ field, (b) the predicted field (from noisy input), (c) the absolute difference
  between the two, and (d) a reference estimation error for noiseless input with 
  identical embedding parameters and training set.
  Panes (e)-(h) and (i)-(l) show the same for their fields $w$ and $s$, respectively.
  The embedding parameters are listed in Table~\ref{tab:bocf_errs}.}
  \label{fig:bocf_u_pred}
\end{figure}

To visualize a few results we chose the best and worst performing estimations. 
Figure~\ref{fig:bocf_w_pred} contains results for $w_{noisy} \rightarrow \{u,v,s\}$ and
Fig.~\ref{fig:bocf_u_pred} shows estimations from a noisy $u$ field to all other variables.
The \text{NRMSE} values in Table~\ref{tab:bocf_errs} indicate that the estimations from field
$w$ perform about one order of magnitude worse than the estimations from field $u$.
Figures~\ref{fig:bocf_w_pred} and \ref{fig:bocf_u_pred} on the other hand reveal
that, even in the latter estimations, the erroneous pixels are concentrated
around the wavefronts. Thus the overall prediction for most of the area is very accurate
in both cases.

\begin{table}
\center
  \begin{tabular}{@{}lllllll}
   \hline
                     &$\gamma$& $\tau$  & $r$  & $D_E$ & $D_R$ & $\text{NRMSE}$\\
    \hline
    $u\rightarrow v$ & 100  & 1 & 1  & 505 &10    & $0.037\pm0.006$ \\
    $u\rightarrow w$ & 100  & 1 & 1  & 505 &15    & $0.048\pm0.007$ \\
    $u\rightarrow s$ & 100  & 1 & 1  & 505 &15    & $0.034\pm0.004$ \\
    $v\rightarrow u$ & 50   & 3 & 2  & 663&25    & $0.093\pm0.015$ \\
    $v\rightarrow w$ & 200  & 1 & 1  & 1005&25    & $0.053\pm0.011$ \\ 
    $v\rightarrow s$ & 200  & 1 & 1  & 1005&25    & $0.081\pm0.015$ \\
    $w\rightarrow u$ & 10   & 2 & 4  & 539&9     & $0.25\pm0.03$   \\
    $w\rightarrow v$ & 10   & 2 & 4  & 539&9     & $0.25\pm0.04$   \\
    $w\rightarrow s$ & 10   & 2 & 4  & 539&9     & $0.21\pm0.03$   \\
    $s\rightarrow u$ & 50   & 1 & 1.5& 459&20    & $0.050\pm0.005$ \\
    $s\rightarrow v$ & 100  & 1 & 1  & 505&15    & $0.060\pm0.009$ \\
    $s\rightarrow w$ & 100  & 1 & 1  & 505&15    & $0.070\pm0.008$ \\
    \hline
    \end{tabular}
     \caption{\label{tab:bocf_errs}
  Embedding parameters and cross estimation errors, averaged over 20 frames, 
  for noisy data from the BOCF model (\ref{eq:bocf}) with temporal sampling of $t_{\text{sample}}=2.0$. 
  A value of 200 was used for the pixels beyond the boundary. $D_E$ is  the original reconstruction dimension and
  $D_R$ is the reduced dimension used for nearest neighbour searches.}
\end{table}


\section{Iterated Time Series Prediction}
In the following we will analyze the performance of local modelling for spatially
extended systems in the context of iterated time series prediction. For this we use
the Kuramoto-Sivashinsky model (\ref{eq:ks}) and the Barkley model (\ref{eq:bk}).

The obvious performance measure in this case is the time it takes before the prediction
errors exceed a certain threshold. Time however is not an absolute concept in dimensionless
systems. Therefore we will also define characteristic timescales of each system
which will give a context to the prediction times.

\subsection{Predicting Barkley Dynamics}
The datasets used during cross estimation were sampled with $t_{\text{sample}}=0.01$ which could be
considered nearly continuous relative to the timescale of the dynamics.
To provide a useful example for temporal prediction 
with a reasonable amount of predicted frames we use a larger 
time step $t_{\text{sample}}=0.2$ (simulation time step was still sufficiently small for accurate numeric integration).

\begin{figure}
  \centering
  \includegraphics[width=0.85\textwidth,keepaspectratio=true]  {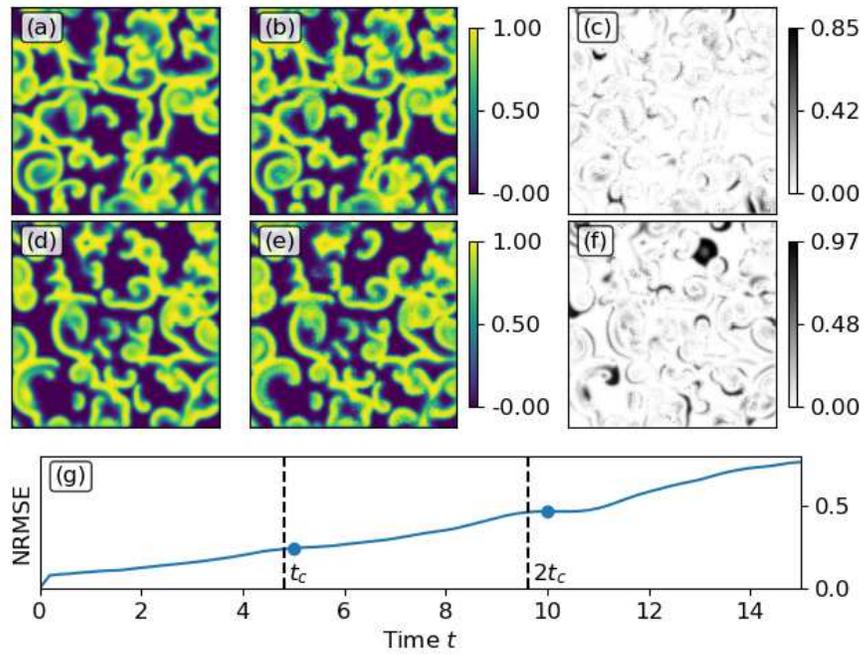}
  \caption{Predicting field $u$ of the Barkley model with system size $150 \times 150$  and training
  of 5000 states. The embedding parameters are $\gamma=12$, $\tau=2$, $r=4$, boundary value $200$.
  PCA reduced the dimension from $D_E = 637$ to $D_R=15$. Panes (a) and (d) show the true evolution at
  time $t=5$ and $t=10$. Panes (b) and (e) contain the iterated prediction at that time and
  (c) and (f) the corresponding absolute error. (g) shows the accumulation of the \text{NRMSE} in the prediction. The dashed lines note $t_c$, the bullets note the times 5 and 10.}
  \label{fig:ks_pred_u}
\end{figure}

Figure~\ref{fig:ks_pred_u} shows one such prediction of the $u$ variable in the
Barkley model.
The figure consists of seven subplots where the top two rows show the system state
at the prediction time steps $n=25,50$ as well as the corresponding iterated predictions.
The very right column displays the absolute errors of the prediction defined by
$|u_{t,\alpha}-\hat u_{t,\alpha}|$.
At the bottom is the time evolution of the $\text{NRMSE}$ for the prediction.
Looking closely at the snapshots in the figure reveals that indeed the maximum prediction error increases quickly, as can be seen by the dark spots of the error plots (c) and (f).
The overall error however increases much more slowly which is confirmed by comparing the
original state with the prediction.

To set the above results into perspective we calculate a characteristic timescale for the
Barkley model. Here we will use the average time between two consecutive local maxima for each pixel, which in good approximation gives the average period of the rotating spiral waves. Averaging over $100\times 100$ pixels and 4000 time steps gave this time as
$t_c = 4.81$.
This means that the error of the $u$ field prediction increased to
$\text{NRMSE}(u, 2t_c)\approx0.5$ within two characteristic times.

\subsection{Predicting Kuramoto-Sivashinsky Dynamics}
The Kuramoto-Sivashinsky (KS) model (\ref{eq:ks}) is a one dimensional system that has just a single field. 
As in the iterated time series prediction of the Barkley model we will need a characteristic
timescale for the dynamics of the KS model to assess the quality of the forecast. The approach of measuring
wave-front-return-times proved not to be as effective. Instead we will use the
Lyapunov time which was defined and calculated for the KS model by Pathak et al. \cite{pathak_model-free_2018}.
The following figures are scaled according to the Lyapunov time $\Lambda t$ with $\Lambda\approx0.1$.

\begin{figure}[h!]

  \includegraphics[width=\textwidth,keepaspectratio=true] {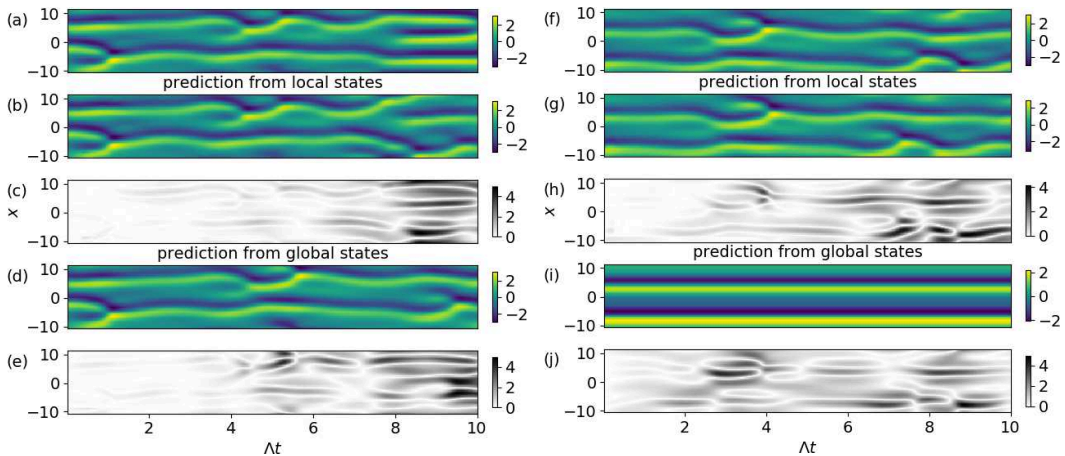}
 \caption{Predictions of the KS dynamics using PCA and  5 nearest neighbours. 
 Shown are:  In (a) and (f) actual evolutions, below it in (b) and (g)
predictions from local states with parameters $\gamma=7,\,\tau=1,\, r=10$,
and at the bottom in (d) and (i) predictions using global states ($\gamma=0,\,r=32$), each along with its errors. 
The left panel was generated using a training set consisting of $9\cdot10^5$ states and
while the right one used $1\cdot10^5$ training states.}
\label{fig:ks_global_local_comparison}
\end{figure}

It is possible to integrate the KS model with different sizes $L$ and spatial samplings $Q$.
We will attempt to predict the time evolution for $L=22$, $Q=64$ and a larger system
with $L=200$ and $Q=512$.
The smaller one of the two has just 64 points and thereby could be predicted by
reconstructing either local or global states, where the latter are given by combining samples from all $Q$ sites in a state vector.
The global states have a higher dimension
and need larger training sets to densely fill the reconstructed space
but in return each vector represents the state of the whole system.
Both approaches are compared in Fig.~\ref{fig:ks_global_local_comparison} using the
same training set of $9\cdot10^5$ states. For smaller training sets
the global state prediction fails altogether as the high dimensional embedding space
is too sparsely sampled and nearest neighbour searches always find the same state over
and over again.
The local state prediction on the other hand still predicts well for some time
as is shown in the same figure.

\begin{figure}[h!]

  \includegraphics[width=\textwidth,keepaspectratio=true] {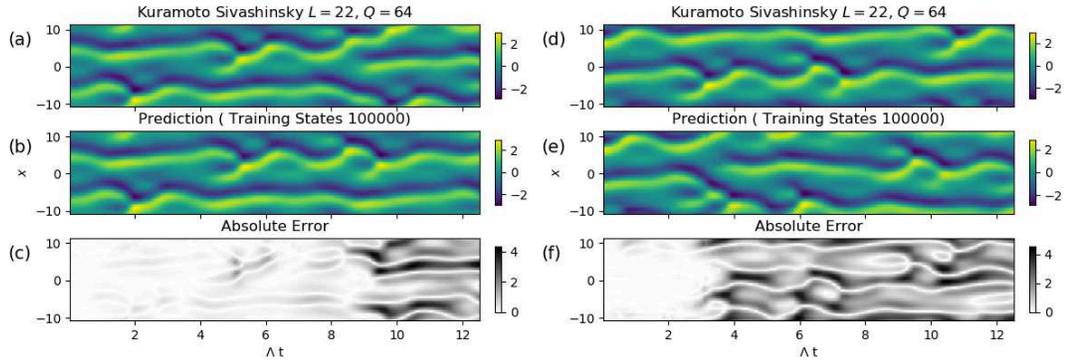}
\caption{Two predictions of the KS dynamics with $L=22,\,Q=64$. All parameters were
identically $\gamma=7,\,\tau=10,\,r=10$ using $10^5$ training states, PCA, and 1 nearest neighbour for local modelling. The only difference lies in the initial condition. (a) and (d)
show the true evolution, (b) and (e) the predictions, and (c) and (f) the
corresponding absolute errors.}
\label{fig:ks_pred_variation}
\end{figure}

A notable observation with the KS model is its variable predictability as it
strongly depends on the initial conditions i.e. the current position on the chaotic attractor. 
Figure~\ref{fig:ks_pred_variation} shows two predictions using $10^5$ training states and identical (reconstruction)
parameters.
The only difference is that the training set chosen from pre-generated data
was offset by 4000 states as compared to the other. The results differ greatly.
In one case the errors stay small for roughly 8 Lyapunov times while the other
diverges already after 3 Lyapunov times.

\begin{figure}[h!]
  \centering
  \includegraphics[width=0.7\textwidth,keepaspectratio=true]   {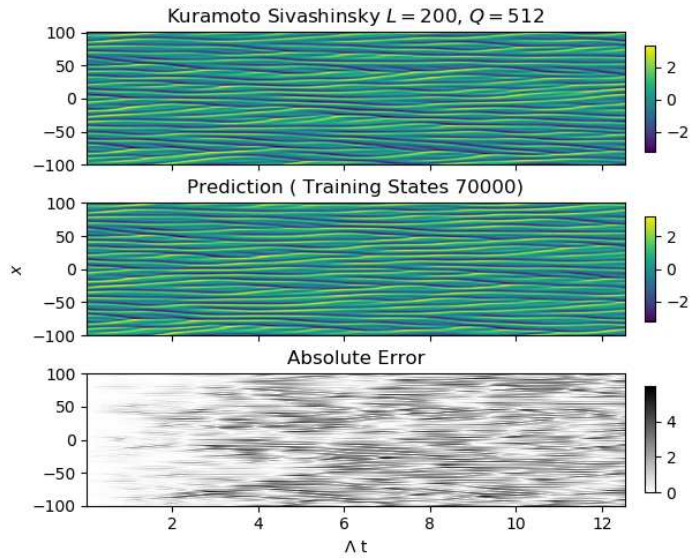}
  \caption{Prediction of the KS dynamics with system parameters $L=200$, $Q=512$ and training
  of 70000 states. The embedding parameters are $\gamma=2$, $\tau=4$, $r=10$ with PCA
  and just one nearest neighbour for local modelling.}
  \label{fig:ks_pred_L200}
\end{figure}

Similar variations in predictability were \emph{not} observed in the KS system with
$L=200$ and $Q=512$, which may be due to the significantly larger extent of the system.
Instead predictions rarely stayed correct for more than two Lyapunov times.
An example is shown in Fig.~\ref{fig:ks_pred_L200}.

The issue of variations in predictability of the KS model
hinder direct comparisons to the work of Pathak et al. \cite{pathak_model-free_2018}
who did not address this problem.
In the small system we saw initial conditions where predictions outperformed
the ones by Pathak et al. but also others that were much worse.
The larger system however has so far been harder to predict and we did
not match the prediction accuracy of the approach of Pathak et al.

\subsection{Benchmark of PCA}
\label{sec:bench}
In this paper we use principal component analysis for two reasons.
The obvious purpose is to find a low-dimensional representation of the high-dimensional embedding.
One very much wanted side-effect is noise reduction. All of the above presented examples used
highly redundant embeddings to allow for noise reduction.

\begin{figure}
    \centering
        \includegraphics[width=\textwidth]  {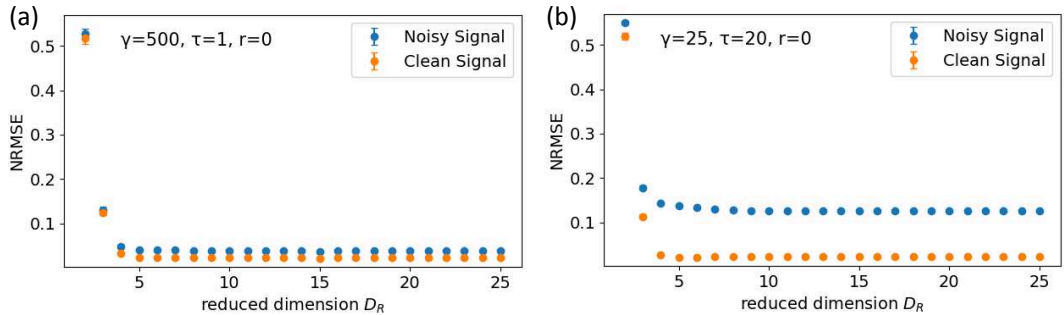}
    \caption{ Normalized Mean Squared Errors of  cross estimation $u\rightarrow v$ of Barkley variables vs. 
    reduced dimension $D_R$   for clean and noisy input signals $u$ with different embedding parameters 
    ((a) $\gamma =500$, $\tau =1$; (b) $\gamma = 25$,   $\tau = 20$)  such that $\gamma \tau$ remains constant.  
    The estimation error increases for very small values of the  reduced dimension $D_R$, but becomes almost 
    constant for $D_R > 5$. For noisy data PCA based dimension reduction based on  higher dimensional  embedding 
    with $D_E=\gamma+1= 501$ in (a)  enables a more efficient noise reduction than the reconstruction with 
     $D_E=\gamma+1 = 26$ in (b).   }
    \label{fig:dimension_dep}
\end{figure}

To evaluate how well PCA is suited for this purpose we test two things:
Does PCA find a low dimensional representation? This is tested in Fig.~\ref{fig:dimension_dep}.
We see the dependence of the prediction error on the output dimension of PCA
in a cross estimation of $u \rightarrow v$ in the Barkley model. 
It is evident that in this case no more than about $5-7$ dimensions are needed to encode all information relevant to
the prediction.

To test whether PCA also successfully eliminated the noise in the test set we compare the two panes in 
Fig.~\ref{fig:dimension_dep} where the results in the right pane were computed
using a 20 times less redundant embedding.
The noiseless predictions perform similarly well in both cases, indicating that the additional
embedding dimensions are indeed redundant and do not add much information to the reconstructed states.
Comparing the noisy predictions highlights the effectiveness of PCA in this case as predictions from the
redundant embedding (Fig.~\ref{fig:dimension_dep}a) are consistently better by one order of magnitude
 (comp. Fig~\ref{fig:dimension_dep}b).

\section{Conclusions}
The combination of local modelling and principal component analysis for dimension reduction
provides a conceptually simple yet effective approach 
to both cross estimation and temporal prediction of complex spatially extended dynamics.
The equations for all three model systems (Barkley model, BOCF model, KS model)
were only needed for data generation
and as such the approach could well be applied to real world data 
where the underlying dynamics are not known.
Adding noise to the input data naturally reduces prediction quality but in
section \ref{sec:bench} it was shown that PCA can restore accuracy from a more redundant embedding.

\section{Acknowlegdements}
The authors thank R. Zimmermann for allowing the use of BOCF model simulation and
S. Luther for scientific discussions and continuous support.

\section{Appendix}
\subsection{Bueno-Orovio-Cherry-Fenton Model}
$H(\cdot)$ denotes the  Heaviside function and the currents $J_{fi},\,J_{so},\, J_{si}$ of the  Buono-Orovio-Cherry-Fenton (BOCF) model (\ref{eq:bocf})  \cite{bueno-orovio_minimal_2008} are defined as:
\begin{align*}
  J_{fi} =& \frac{-v}{\tau_{fi}} H(u-\theta_v)(u-\theta_v)(u_u-u)\\
  J_{so} =& \frac{1}{\tau_o}(u-u_o)(1-H(u-\theta_w)) + \frac{1}{\tau_{so}}H(u-\theta_w)\\
  J_{si} =& \frac{-1}{\tau_{si}}H(u-\theta_w)ws .
\end{align*}
The $\tau$ parameters used above are not constant but rather a function of the cell membrane voltage
variable $u$:
\begin{align*}
  \tau_v^- =& (1-H(u-\theta_v^-))\tau_{v1}^- +H(u-\theta_v^-)\tau_{v2}^-\\
  \tau_w^- =& \tau_{w1}^- + \frac{1}{2}(\tau_{w2}^- -\tau_{w1}^-)(1+\tanh(k_w^-(u-u_w^-)))\\
  \tau_{so}^- =& \tau_{so1}+\frac{1}{2}(\tau_{so2}-\tau_{so1})(1+\tanh(k_{so}(u-u_{so})))\\
  \tau_s =& (1-H(u-\theta_w))\tau_{s1} + H(u-\theta_w)\tau_{s2}\\
  \tau_o =& (1-H(u-\theta_o))\tau_{o1} + H(u-\theta_o)\tau_{o2}\\
\intertext{and}
  v_\infty =& \begin{cases}
    1, u < \theta_v^-\\
    0, u \geq \theta_v^-
\end{cases}\\
w_\infty =& (1-H(u-\theta_o))\left(1-\frac{u}{\tau_{w\infty}}\right) + H(u-\theta_o)w_\infty^*.
\end{align*}
All other parameters are listed in Table~\ref{tab:tnnp}.

\begin{table}
  \parbox{0.24\linewidth}{
    \begin{tabular}{c|c}
    $u_o$         & 0\\
    $u_u$         & 1.58\\
    $\theta_v$    & 0.3\\
    $\theta_w$    & 0.015\\
    $\theta_v^-$  & 0.015\\
    $\theta_o$    & 0.006\\
    $\tau_{v1}^-$ & 60 \\
  \end{tabular}
  }
  \parbox{0.24\linewidth}{
    \begin{tabular}{c|c}
    $\tau_{v2}^-$ & 1150\\
    $\tau_v^+$    & 1.4506\\
    $\tau_{w1}^-$ & 70\\
    $\tau_{w2}^-$ & 20\\
    $k_w^-$       & 65\\
    $u_w^-$       & 0.03\\
    $\tau_w^+$    & 280 \\
  \end{tabular}
  }
  \parbox{0.24\linewidth}{
    \begin{tabular}{c|c}
      $\tau_{fi}$   & 0.11\\
      $\tau_{o1}$   & 6\\
      $\tau_{o2}$   & 6\\
      $\tau_{so1}$  & 43\\
      $\tau_{so2}$  & 0.2\\
      $k_{so}$      & 2\\
      $u_{so}$      & 0.65\\
    \end{tabular}
  }
  \parbox{0.24\linewidth}{
    \begin{tabular}{c|c}
      $\tau_{s1}$   & 2.7342\\
      $\tau_{s2}$   & 3\\
      $k_s$         & 2.0994\\
      $u_s$         & 0.9087\\
      $\tau_{si}$   & 2.8723\\
      $\tau_{w\infty}$ & 0.07\\
      $w_\infty^*$  & 0.94
    \end{tabular}
  }
\caption{\label{tab:tnnp}Parameter set for the BOCF model \cite{bueno-orovio_minimal_2008}
  that imitates the  Ten Tusscher-Noble-Noble-Panfilov model \cite{ten_tusscher_model_2004}.}
\end{table}

\bibliographystyle{spmpsci}      
\bibliography{IDP_references}   

\end{document}